\documentclass{icrc2009}

\usepackage{graphicx}   
\usepackage[caption=false]{caption}    
\usepackage[font=footnotesize]{subfig} 
\usepackage{fixltx2e}
\usepackage{url}

\newcommand{\shorttitle}[1]%
{\markboth{Proceedings of the 31\MakeLowercase{$^{st}$} ICRC, {\L}\'{o}d\'{z} 2009}{#1} }

\usepackage{multirow}
\usepackage{amssymb,amsmath,wrapfig}

\newcommand{\bea}{\begin{eqnarray}}
\newcommand{\eea}{\end{eqnarray}}
\newcommand{\Xmax}{$X_\mathrm{max}$}

\def \gcm   {g~cm$^{-2}$}
\def\Offline{\mbox{$\overline{\textrm%
{Off}}$\hspace{.05em}\protect\raisebox{.4ex}%
{$\protect\underline{\textrm{line}}$}}}
\def \vap {\mathrm{H}_2\mathrm{O}}
\def\eV{\ifmmode {\mathrm{\ e\kern -0.1em V}}\else
                   \textrm{e\kern -0.1em V}\fi}%

\begin{document}
\title{Fluorescence emission induced by extensive air showers in dependence 
on atmospheric conditions}

\author{\IEEEauthorblockN{Bianca Keilhauer\IEEEauthorrefmark{1},
			  Michael Unger\IEEEauthorrefmark{1}}
                            \\
\IEEEauthorblockA{\IEEEauthorrefmark{1}Karlsruhe Institute of Technology (KIT), \\
Forschungszentrum Karlsruhe, Institut f\"ur Kernphysik, P.O.Box 3640, 76021 
Karlsruhe, Germany}}
\shorttitle{Keilhauer, Unger; atmosphere-dependent fluorescence}
\maketitle

\begin{abstract}
Charged particles of extensive air showers (EAS), mainly electrons and positrons, initiate
the emission of fluorescence light in the Earth's atmosphere. This light provides a
calorimetric measurement of the energy of cosmic rays. For reconstructing the primary
energy from an observed light track of an EAS, the fluorescence yield in air has to be
known in dependence on atmospheric conditions, like air temperature, pressure, and
humidity. Several experiments on fluorescence emission have published various sets of data
covering different parts of the dependence of the fluorescence yield on atmospheric
conditions.

Using a compilation of published measurements, a calculation of the fluorescence yield in
dependence on altitude is presented. The fluorescence calculation is applied to simulated
air showers and different atmospheric profiles to estimate the influence of the
atmospheric conditions on the reconstructed shower parameters.
  \end{abstract}


\section{Introduction}\label{sec:Introduction}

The number of
emitted fluorescence photons at the air shower can be written as 
\bea
\frac{d^2N^0_\gamma}{dX d\lambda} = Y(\lambda,P,T,e)\cdot
\frac{dE^\mathrm{tot}_\mathrm{dep}}{dX}, 
\eea 
where $Y(\lambda,P,T,e)$ is the fluorescence yield in dependence on
wavelength $\lambda$, air pressure $P$, air temperature $T$, and vapour 
pressure $e$. The deposited energy of the secondary particles is
denoted as $dE^\mathrm{tot}_\mathrm{dep}/dX$.

In the last couple of years, a lot of effort has been put on the
investigation of atmospheric dependences on nitrogen fluorescence in
air~\cite{NIMA597}. The fluorescence yield $Y_\lambda$ can be written
as
\bea
Y_\lambda = \Phi^0_\lambda \cdot \lambda/hc \cdot \frac{1}{1+P/P^\prime_v},
\label{eq:yield}
\eea
where $\Phi^0_\lambda$ is the fluorescence efficiency at zero
pressure, $P$ is the air pressure, and $P^\prime$ is the
characteristic pressure for which the probability of collisional
quenching equals that of radiative de-excitation. The index $v$ indicates 
the vibrational level of the exited state. Several groups have
already investigated aspects of the fluorescence emission from
nitrogen molecules in air (e.g.\ Bunner~\cite{bunner}, Davidson \&
O'Neil~\cite{DON}, Kakimoto et al.~\cite{kakimoto}, MACFLY~\cite{macfly} and
FLASH~\cite{flash}). In addition there are various ongoing
experimental activities,
e.g.\ AIRFLY~\cite{airfly_Pdep,airfly_Edep,airfly_THumdep,airfly_calib},
Nagano \& Sakaki et al.~\cite{nagano2004,sakaki},
AirLight~\cite{airlight} and Ulrich \& Morozov et al.~\cite{ulrich}.
One major goal of all experiments is to obtain an absolute
fluorescence yield $Y^0_\lambda = \Phi^0_\lambda \cdot \lambda/hc$
either for the main contributing band at 337.1~nm or for the entire
spectrum in the range of interest between about 300 --
420~nm. $Y^0_\lambda$ represents the intrinsic radiative de-excitation
of the nitrogen molecules. However, in gas like air quenching
processes have to be taken into account because the rate of radiative
de-excitations is reduced by collisions between excited nitrogen
molecules and further molecules in the gas. These quenching processes
depend on atmospheric conditions and are described by $(1+P/P^\prime_v)^{-1}$ in
Eq.~(\ref{eq:yield}). Accounting all currently known effects, we can
write
\begin{equation}
\begin{split}
& \frac{P}{P^\prime_v} = \frac{\tau_{0,v}\cdot P_\mathrm{air}\cdot N_A}{R\cdot
  T_\mathrm{air}}\cdot \sqrt{\frac{k\cdot T_\mathrm{air}\cdot N_A}{\pi}} \\ & \cdot
\biggl( 4 C_\mathrm{vol}(\mathrm{N}_2) \cdot \sigma_{\mathrm{NN},v}(T)
\sqrt{M_\mathrm{N}^{-1}} \\ & + 2 C_\mathrm{vol}(\mathrm{O}_2) \cdot
\sigma_{\mathrm{NO},v}(T)
\sqrt{2\Bigl(M_\mathrm{N}^{-1}+M_\mathrm{O}^{-1}\Bigr)} \\ & + 2
C_\mathrm{vol}(\vap) \cdot \sigma^0_{\mathrm{N}\vap,v} 
\sqrt{2\Bigl(M_\mathrm{N}^{-1}+M_{\vap}^{-1}\Bigr)}\biggr),
\label{eq:ppprime}
\end{split}
\end{equation}
with $\tau_{0,v}$ as the mean life time of the radiative transition to
any lower state, the index $v$ indicates again the vibrational level of the 
exited state as for $P^\prime_v$, $N_A$ is Avogadro's number, $R$ is the universal gas
constant, $T_\mathrm{air}$ is the air temperature, $k$ is the Boltzmann
constant, $C_\mathrm{vol}$ is the fractional part per volume of the relevant gas
constituents, and $M_x$ is the mass per mole where $x$ stands for
the relevant gas constituents. Up to now, the collisional cross sections
$\sigma_{\mathrm{N}x,v}$ have been taken as temperature-independent
even though it was known from theory that there has to be a
temperature dependence. Recently, first experiments could confirm this
dependence for nitrogen-nitrogen and nitrogen-oxygen quenching. 
The temperature-dependence of the nitrogen-vapour quenching has not been
measured yet. First estimates indicate only minor importance with an
effect of less than 1\% change in the reconstructed energy of an air
shower~\cite{ulrich2009}. An independent measurement of the 
temperature-dependent collisional cross sections in air has been performed
quite recently. First analyses of data indicate compatible results with
the measurements from AIRFLY and will be published soon~\cite{ulrich2009}. 

Adopting this description of fluorescence emission for air shower
reconstruction, we have to apply atmospheric profiles
for temperature, pressure, and vapour pressure. This cannot be provided by
simple atmospheric models as these usually do not include vapour profiles. 
However, profiles obtained with meteorological radio soundings do provide all 
necessary quantities~\cite{soundings}. 

\section{Fluorescence Models in Reconstruction}
\label{sec:reco}

For this study, we could use the simulation and reconstruction framework 
\Offline~\cite{offline}
of the Pierre Auger Observatory~\cite{pao}. Within this framework, we could
obtain standard monthly models for the area of that observatory which do not
include water vapour profiles~\cite{soundings}. Additionally, we had access to 
109 actual nightly atmospheric profiles from local radio soundings that cover 
all conditions within a year. One of the advantages of the framework is that it
features many implementations of different fluorescence models which can easily
be interchanged.  

The first implementation of a fluorescence model in \Offline, referred to as K96, 
is based on measurements by Kakimoto et
al.~\cite{kakimoto}. The fluorescence yield is parametrised in
dependence on deposited energy and on altitude by considering the
pressure and $\sqrt{T}$-dependences.
The second fluorescence model, N04, has the same functional form of
parametrisation and describes data from Nagano et
al.~\cite{nagano2003,nagano2004}. These measurements provide
spectrally resolved data for 15 wavelengths between 300 and
430~nm. Also in this description, only the pressure and
$\sqrt{T}$-dependences are considered.
The third fluorescence description in \Offline \ is given by the AIRFLY
Collaboration in 2007, labelled with A07. The fluorescence yield is given
as~\cite{airfly_THumdep}
\bea 
Y_\lambda(P,T) = Y^{337}_{P_0,T_0} \cdot I^\lambda_{P_0,T_0} \cdot
\frac{1+\frac{P_0}{P^\prime(\lambda,T_0)}}{1+\frac{P_0}{P^\prime(\lambda,T_0)\sqrt{T/T_0}}}.
\eea
$Y^{337}_{P_0,T_0}$ is the fluorescence yield at 337.1~nm as measured
at their standard experimental conditions which are $P_0$ = 800~hPa
and $T_0$ = 293~K. The other transitions have been measured relatively
to that at 337.1~nm and are given by $I^\lambda_{P_0,T_0}$. Overall, 34
transitions could be resolved between 295 and 430~nm. 
Since the 
absolute calibration of this experiment is still under
study, $Y_{337}$ is 
normalised to the corresponding value of N04. It should be
pointed out that the description in this model can easily be expanded
to account for vapour quenching and temperature-dependent
collisional cross sections.
The fourth implementation of a fluorescence model follows the
calculation from Keilhauer et al.\ in 2008~\cite{NIMA99}. Here, 23
wavelengths between 300 and 430~nm are considered by applying
Eq.~(\ref{eq:yield}) and (\ref{eq:ppprime}). The model uses a compilation
of different measurements~\cite{NIMA99,keilhauer2006}. For the temperature-dependent 
collisional cross
sections, the data from AIRFLY~\cite{airfly_THumdep} are used. These
$\alpha$-coefficients are obtained in air, so the same
$\alpha_\lambda$ is applied to NN-collisions and
NO-collisions. The temperature-dependent collisional cross sections in Eq.~(\ref{eq:ppprime})
are written as $\sigma_{\text{N}x,\nu}(T) = \sigma^0_{\text{N}x,\nu}\cdot 
T^{\alpha_\nu}$ where $\sigma^0_{\text{N}x,\nu} = \sigma_{\text{N}x,\nu}\cdot 
293^{-\alpha_\nu}$ is the measured temperature-independent cross section at 
standard experimental conditions of $T$ = 293~K.
Cross sections for nitrogen - water
vapour collisions have been measured by two
experiments~\cite{morozov,waldenmaier}.

\section{Atmosphere-dependent Fluorescence Emission}
\label{sec:analysis}

To study the overall effect of different fluorescence models on
reconstructed air shower observables, primary energy $E$ and 
position of shower maximum \Xmax, it is
important to account for only that part of the fluorescence spectrum
that a detector is sensitive to as well as the wavelength dependent
attenuation in the atmosphere (see for instance Fig.~8d
in~\cite{nagano2004}). Moreover, since the atmospheric parameters $P$,
$e$ and $T$ depend on altitude, different fluorescence
models will propagate differently to $E$ and \Xmax \ if the
shower reached its maximum high in the atmosphere or close to the
ground.

To include all these effects, we proceeded as follows: 
Proton and iron showers with energies between
10$^{17.5}$ and 10$^{20}$ \eV \ were generated using
{\scshape Conex}~\cite{Bergmann:2006yz} and
{\scshape QGSJETII}~\cite{Ostapchenko:2004ss}.
The fluorescence light was generated according to the K08
model including water vapour quenching and 
temperature-dependent collisional cross sections. The events were generated
with time stamps that corresponds to nights with balloon launches, such
that realistic profiles for $P$, $e$ and $T$ could be obtained.
In the following, we will compare the difference in the reconstructed
$E$ and \Xmax \ values of these simulated showers.

\subsection{Fluorescence Models}
\label{sec:flmodels}

\begin{figure*}[htbp]
    \begin{minipage}[c]{0.23\textwidth}
  \caption{Comparison of the influence of different fluorescence models 
           on $E$ and \Xmax \ (without vapour quenching and
           temperature-dependent collisional cross sections). The abbreviations 
           of the different fluorescence models are defined in 
           Sec.~\ref{sec:reco}.}
        \label{fig:plainModels}
    \end{minipage} \hfill
    \begin{minipage}[c]{0.75\textwidth}
     \includegraphics[width=\linewidth]{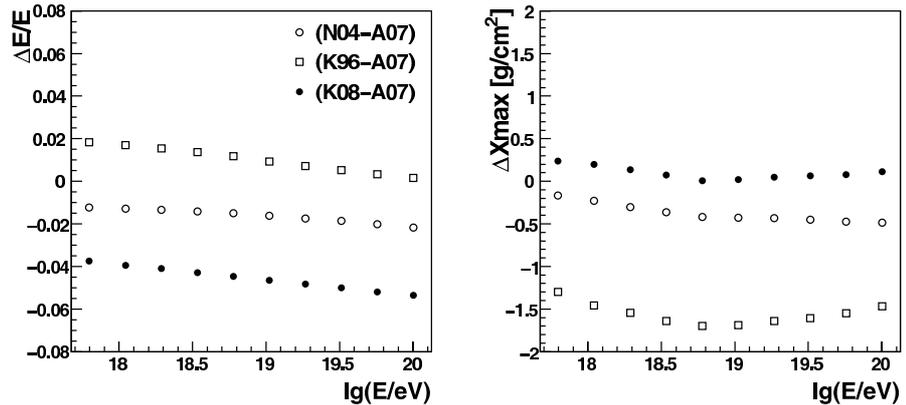}
    \end{minipage}
\vspace{-24pt}
\end{figure*}

The \Xmax \ and energy differences for reconstructions with different
fluorescence models is shown in Fig.~\ref{fig:plainModels}.  
For this
figure, the water vapour quenching and temperature-dependent
collisional cross sections were {\itshape not} switched on in the
K08-model, thus this comparison is only sensitive to the $Y(P,T)$
implementations. As explained above, the A07 model is
normalised to N04, therefore they are not independent and show
correspondingly the smallest differences.

\subsection{Temperature-dependent collisional cross sections and vapour quenching}
\label{sec:dependences}

\begin{figure*}[htbp]
    \begin{minipage}[c]{0.23\textwidth}  
     \caption[k08comp]{Comparison of the effect of switching off $\sigma_{e}$ 
           and the collisional cross sections $\sigma_{T}$ on $E$ and \Xmax \
           as well as the influence of different
           vapour quenching, $\sigma_{e}^{W}$~\cite{waldenmaier} 
           and $\sigma_{e}^{M}$~\cite{morozov}.}
        \label{fig:k08Comp}
    \end{minipage} \hfill
    \begin{minipage}[c]{0.75\textwidth}
    \includegraphics[width=\linewidth]{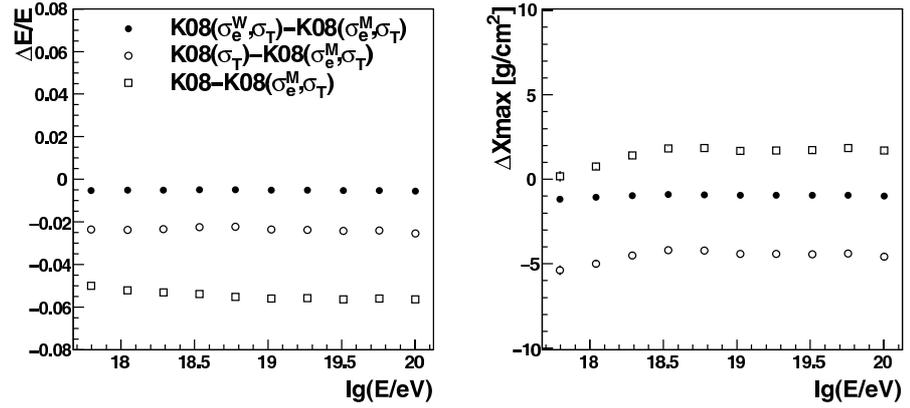}
    \end{minipage}
\vspace{-12pt}
\end{figure*}

The influence of the water vapour quenching ($\sigma_e$) and
temperature-dependent collisional cross sections ($\sigma_T$) on
\Xmax \ and $E$ was studied by subsequently switching off the effects
in the reconstruction using the K08-model. As can be seen in the left panel of
Fig.~\ref{fig:k08Comp}, ignoring $\sigma_e$- and $\sigma_T$-effects leads to an
underestimation of the reconstructed energy by about 5\%.  Both
$\sigma_e$- and $\sigma_T$-dependences affect the shape of the longitudinal
profile. Since the $\sigma_e$-dependence is most important close to ground and
the $\sigma_T$-dependence affects mainly higher altitudes, the two effects partially
compensate (see right panel of Fig.~\ref{fig:k08Comp}) 
leading to only a small \Xmax \ shift of $\le$ 2~\gcm.

Interchanging the water vapour quenching from~\cite{morozov} with the
independent measurement from~\cite{waldenmaier} affects the shower
observables very little (see solid black dots in
Fig.~\ref{fig:k08Comp}).

\begin{figure*}[htbp]
    \begin{minipage}[c]{0.23\textwidth} 
 \caption{Difference in reconstructed energy and \Xmax \ in dependence the vertical
           height of the shower maximum ($E=10^{19}$ \eV).}
        \label{fig:deltaEvsXmax} 
    \end{minipage} \hfill
    \begin{minipage}[c]{0.75\textwidth}
\includegraphics[width=\linewidth]{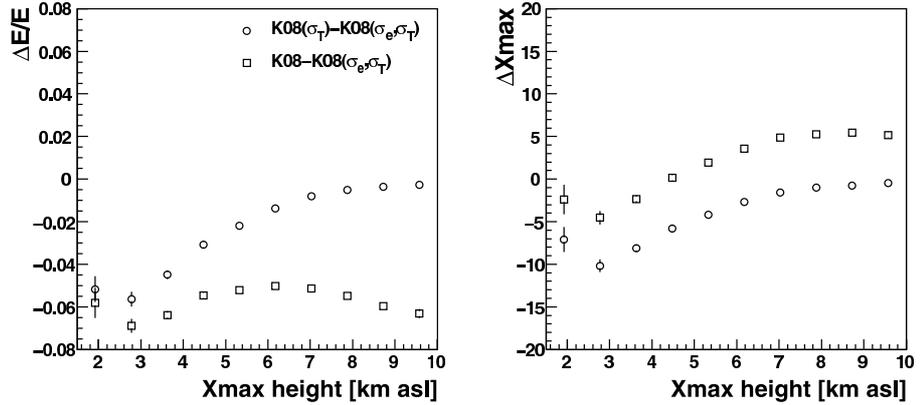}
    \end{minipage}
\vspace{-12pt}
\end{figure*}

The varying strengths of the $\sigma_e$- and $\sigma_T$-dependences at different altitudes
can be seen in Fig.~\ref{fig:deltaEvsXmax}.
Ignoring the $\sigma_T$-effect, the energy is misreconstructed up to -7\% for
showers with \Xmax \ high up in the atmosphere. Ignoring the
$\sigma_e$-dependence, the energy is underestimated also up to 7\% for 
showers with \Xmax \ close to ground. The position of shower maximum is also 
affected with the largest biases being observed for deep and shallow
showers.  The overall shift of \Xmax \ is
strongest for showers with a position of shower maximum at about 3~km~a.s.l.\ with -5~\gcm
\, or for showers with \Xmax \ at 9~km~a.s.l.\ with 5~\gcm. It can clearly be seen in the
right-hand plot of Fig.~\ref{fig:deltaEvsXmax} that the $\sigma_e$-dependence cancels out
partly the $\sigma_T$-dependence concerning \Xmax.

\section{Discussion of Results}
\label{sec:uncertainties}

In the fluorescence model K08, all currently known effects of the fluorescence light 
emission are included in dependence on varying atmospheric conditions. Running this 
model in combination with actual atmospheric profiles, gives a good estimate of the overall mis-reconstruction and uncertainties 
of a standard reconstruction. However, it must be stressed that all of the models used in
this study have a reported uncertainty of the absolute fluorescence yield well above 10\%.
In particular, the AIRFLY and AirLight experiments will perform an absolute 
fluorescence yield calibration with higher accuracy and results can be expected within one
year. 

\begin{figure*}[htbp]
    \begin{minipage}[c]{0.23\textwidth} 
        \caption[k08a07]{Difference of the reconstruction using 
                 the full K08 model to A07. Error
                 bars denote the RMS spread. Note that we corrected for the 'trivial' 
                 yield difference, $\Delta$, from Fig.~\ref{fig:plainModels}.}
        \label{fig:k08a07Comp}
    \end{minipage} \hfill
    \begin{minipage}[c]{0.75\textwidth}
 \includegraphics[width=\linewidth] {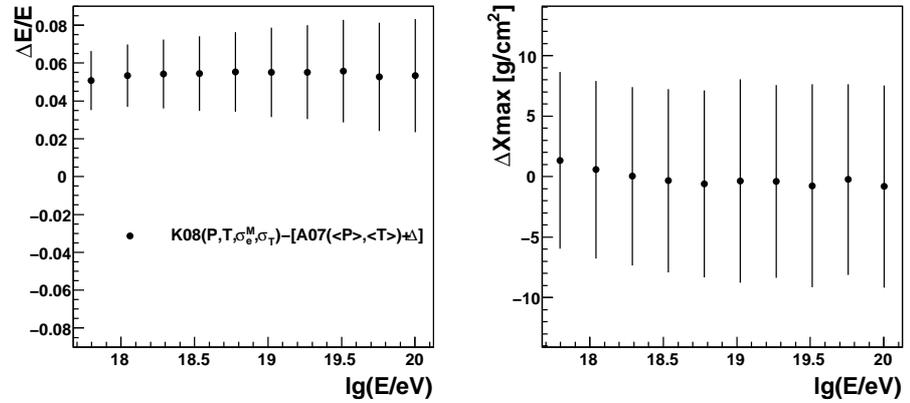}
    \end{minipage}
\vspace{-12pt}
\end{figure*}

In Fig.~\ref{fig:k08a07Comp}, the difference of the reconstruction of $E$ and \Xmax \
using the K08 model with all effects in combination with actual atmospheric profiles and
a standard reconstruction with the A07 fluorescence model with monthly models 
can be seen. More or less independent of energy, the reconstructed 
primary energy $E$ is higher by about 5\% using K08 compared with A07 model. The position
of shower maximum \Xmax \ is nearly unaffected. These results are very similar to the
comparison of the full K08 model and that without $\sigma_e$- and $\sigma_T$-dependences.
Thus, no additional systematics are introduced while changing the fluorescence model apart 
from those obtained by the $\sigma_e$- and $\sigma_T$-dependences.

Studying the variation in $E$ and \Xmax \ in dependence on the
height of the shower maximum, two
extreme cases can be found: The average shift 
in $E$ can be up to -7\% for $E$ and -5~\gcm \, for \Xmax \ for deeply-penetrating showers
and up to -7\% for $E$ and +5~\gcm \, for \Xmax \ for showers that develop
high in the atmosphere.

Furthermore, we studied the influence of different types of primary
particle in terms of proton- and iron-induced showers. 
Comparing the widths of the distribution, no difference could be 
found between proton- and iron-induced air showers.

The change in the atmosphere description from monthly models to actual sounding
profiles do hardly affect the reconstructed energy nor the position of shower maximum. For
$E$, the difference is well below 1\% and for \Xmax \ below 2~\gcm.

Obviously, the fluctuation of the atmosphere around the monthly
average atmosphere values adds an additional contribution to the
statistical uncertainty of the reconstructed energy and \Xmax \ of one
shower. The 'end-to-end' comparison of the A07 model with monthly
averages to the K08 model with sounding data yields RMS$(\Delta E/E)\in
[1.5,3.0]$\% and RMS(\Xmax)$~ \in [7.2,8.4]$ \gcm (cf.\ Fig.~\ref{fig:k08a07Comp}).

Finally, the systematic difference in the collisional cross section data from two
independent measurements~\cite{morozov,waldenmaier} are negligible. The reconstructed
energy varies less than 1\% and the position of shower maximum about 1~\gcm \, while 
interchanging the cross sections. Varying the $\alpha$-coefficients for the
temperature-dependent collisional cross sections within their given uncertainties, yields
in less than 1\% change in reconstructed energy as well.

\section*{Acknowledgements}

The authors would like to thank the Pierre Auger Collaboration for
providing the simulation and reconstruction framework used in this work. Part of this 
work is supported by the BMBF under contract 05A08VK1.

\end{document}